\documentclass{PoS}

\input phys_defs

\title{Measurements of the Partial Branching Fraction for $\Bxulnu$ and the Determination of $\Vub$}

\ShortTitle{Determination of \Vub using $\Delta {\cal{B}}(\Bxulnu)$}

\author{\speaker{Michael Sigamani}\thanks{On behalf of the $\babar$ collaboration.}\\
        INFN sezione di Universit\`{a} di Roma `La Sapienza'\\
        E-mail: \email{sigamani@cern.ch}}

\abstract{
We measure partial branching fractions for the inclusive charmless semileptonic
decay of $\Bxulnu$, and present determinations of the CKM matrix element $\Vub$.
This analysis is based on a sample of 467 million $\FourS$ decays into $\BB$
pairs, collected with the \babar\ detector at the PEP-II $e^{+} e^{-}$ storage rings.
We select events where one of the B mesons is fully reconstructed in a hadronic decay mode and the other $B$
decays semileptonically into an electron or a muon. We then use the invariant mass, $\mx$, of the hadronic system,
the invariant mass squared, $q^{2}$, of the lepton and neutrino pair, the lepton momentum
in the in the $B$ meson rest frame $\Pl$, the kinematic variable $P_{+}$ or one of their combinations
as discriminating variables to isolate $\Bxulnu$ decays from the background.
We then measure the following partial branching fractions in various regions of phase space:
$\Delta {\cal{B}} =(1.08 \pm 0.08_{\rm stat.} \pm 0.06_{\rm sys.} \pm 0.01_{\rm theo.}) \times 10^{-3}$
($M_{X}<1.55~{\rm GeV}/c^{2},\Pl>1.0 ~GeV/c$),
$\Delta {\cal{B}} =(1.15 \pm 0.10_{\rm stat.} \pm 0.08_{\rm sys.} \pm 0.01_{\rm theo.}) \times 10^{-3}$
($M_{X}<1.70~{\rm GeV}/c^{2},\Pl>1.0 ~GeV/c$),
$\Delta {\cal{B}} =(0.98 \pm 0.09_{\rm stat.} \pm 0.08_{\rm sys.} \pm 0.01_{\rm theo.}) \times 10^{-3}$
($P_{+} < 0.66~{\rm GeV},\Pl>1.0 ~GeV/c$),
$\Delta {\cal{B}} =(0.68 \pm 0.06_{\rm stat.} \pm 0.04_{\rm sys.} \pm 0.02_{\rm theo.}) \times 10^{-3}$
($M_{X}<1.7~{\rm GeV}/c^{2}$, $q^{2}>8~{\rm GeV}^{2}/c^{4},\Pl>1.0 ~GeV/c$),
$\Delta {\cal{B}} =(1.50 \pm 0.13_{\rm stat.} \pm 0.14_{\rm sys.} \pm 0.02_{\rm theo.}) \times 10^{-3}$
($\Pl>1.3 ~GeV/c$).
These partial branching fractions are translated into values of $\Vub$ using several theoretical calculations.
An estimate based on the most inclusive analysis gives $\Vub = (4.31\pm 0.35) \times 10^{-3}$.
Furthermore, we place a limit on isospin breaking in $B\to X_u\ell\nu$ decays of 9\% at 90\% C.L.
using separate measurements of partial branching ratios for $B^0$ and $B^+$.

}

\FullConference{35th International Conference of High Energy Physics - ICHEP2010,\\
                July 22-28, 2010\\
                Paris France}

\begin{document}

\section{Introduction}

In the Standard Model of electroweak interactions,
the coupling strength of the $b$ to the $u$ quark in
the weak interaction is described by the Cabibbo-Kobayashi-Maskawa
(CKM)~\cite{Kobayashi:1973fv} matrix element $\Vub$.
A precise determination of $\Vub$ is therefore crucial for probing
the CKM mechanism for quark mixing.
Experimentally, $\Vub$ is obtained using branching fraction measurements of charmless
$\Bxulnu$ decays. However, separating $\Bxulnu$ from the $\Bxclnu$ background,
whose rate is some fifty times larger is very difficult due to the fact that it
shares a very similar event topology as the signal.
Therefore, for inclusive measurements we are forced to restrict the experimental phase space to regions where
this background is highly suppressed and measure partial branching fractions (PBFs).
We extrapolate back to the full phase space using $QCD$-based theoretical
calculations that are themselves affected by uncertainties which increase if the phase space is reduced.
We then try to find a balance between the theoretical and experimental uncertainties in selecting
regions which minimise the total uncertainty on $\Vub$.

\section{Analysis technique}

\begin{figure}[htb]
  \begin{center}
    \includegraphics[width=0.48\textwidth]{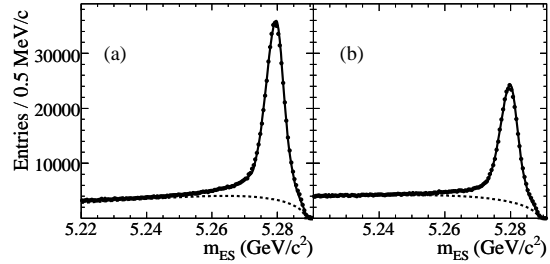}
    \caption{Fit of \mes~distributions of
     $\Bxulnu$
     candidates passing the semileptonic selection criteria in MC events
     where a \Bpm (a) or a \Bz (b) has been fully reconstructed in a hadronic decay mode.}
    \label{fig:mes:pstarsample}
  \end{center}
\end{figure}

We begin by selecting events whereby one of the $B$ mesons from the $\FourS$ decay is fully
reconstructed in a hadronic decay mode ($B_{reco}$).
We then identify a semileptonic decay by requiring an electron or muon from the opposite $B$ meson ($B_{recoil}$) in the event.
This technique leaves a clean $B_{recoil}$ sample with both the flavour and four-momentum of the event well determined.
The remaining particles are associated to the hadronic system $X$,
and the detection of missing energy is associated to the neutrino.
We use an un-binned maximum likelihood fit to the \mes~
\footnote{$\mes = \sqrt{s/4 - \vec{p}^{\,2}_B}$, where $\sqrt{s}$ is the total energy of the \FourS and
and ${p}_B = $ is the momentum of the $\breco$.} distribution
in order to suppress the combinatorial background (badly reconstructed $B$ mesons),
and the continuum background ($\epem \to \qqbar$).
We use an ARGUS \cite{argusf} function to describe the sum of both the combinatorial and continuum background,
and a modified Crystal-Ball function \cite{cristal-ball} to describe the signal peak. The results are shown in
Fig.~\ref{fig:mes:pstarsample}.

In our resulting $B_{recoil}$ sample we are left with various physics processes including signal
$\Bxulnu$ and $\Bxclnu$ semileptonic decays.
All other processes are referred to as ``other''.
The backgrounds include leptons found in secondary cascade $B\to D\to \ell$ or $B\to D_s\to \tau$ decays,
leptons from $\tau$ decays, leptons from $\jpsi\to \ell\ell$,
or hadrons mis-identified as leptons (mainly affects the muon sample).
In order to suppress these backgrounds we perform various selection cuts on the $B_{recoil}$.
Firstly, we require one lepton with momentum in the $B_{recoil}$ rest-frame of $\Pl>1\gevc$.
This reduces the charm background since secondary leptons are frequent in
$\Bxclnu$ due to cascade decays of the $X_c$. Whereas, secondary leptons are very rare
in $\Bxulnu$ decays.
Also, we require $Q_{B_{recoil}} Q_\ell<0$, where $Q_{B_{recoil}}$ is the charge of the
$B_{recoil}$ and $Q_\ell$ is the charge of the primary lepton.
Furthermore, we set $Q_{tot}=Q_{B_{reco}}+Q_{X}+Q_{\ell}=0$, where $Q_{B_{reco}}$ is the charge of the
$B_{reco}$ and $Q_{X}$ is the charge of the hadronic system.
Also, we reduce $B\to D^* \ell\nu$ decays by exploiting the properties of the soft pion produced in the $D^*\to D\pi$ decay
in order to infer and place cuts on the missing mass.
The only undetected particle should be a neutrino, so we require the missing mass squared to be less than $0.5 \gevccsq$.
Lastly, we veto charged kaons and $\KS$ found in the $B_{recoil}$.

\begin{figure*}[t]
    \begin{centering}
      \includegraphics[width=0.245\textwidth,totalheight=7cm]{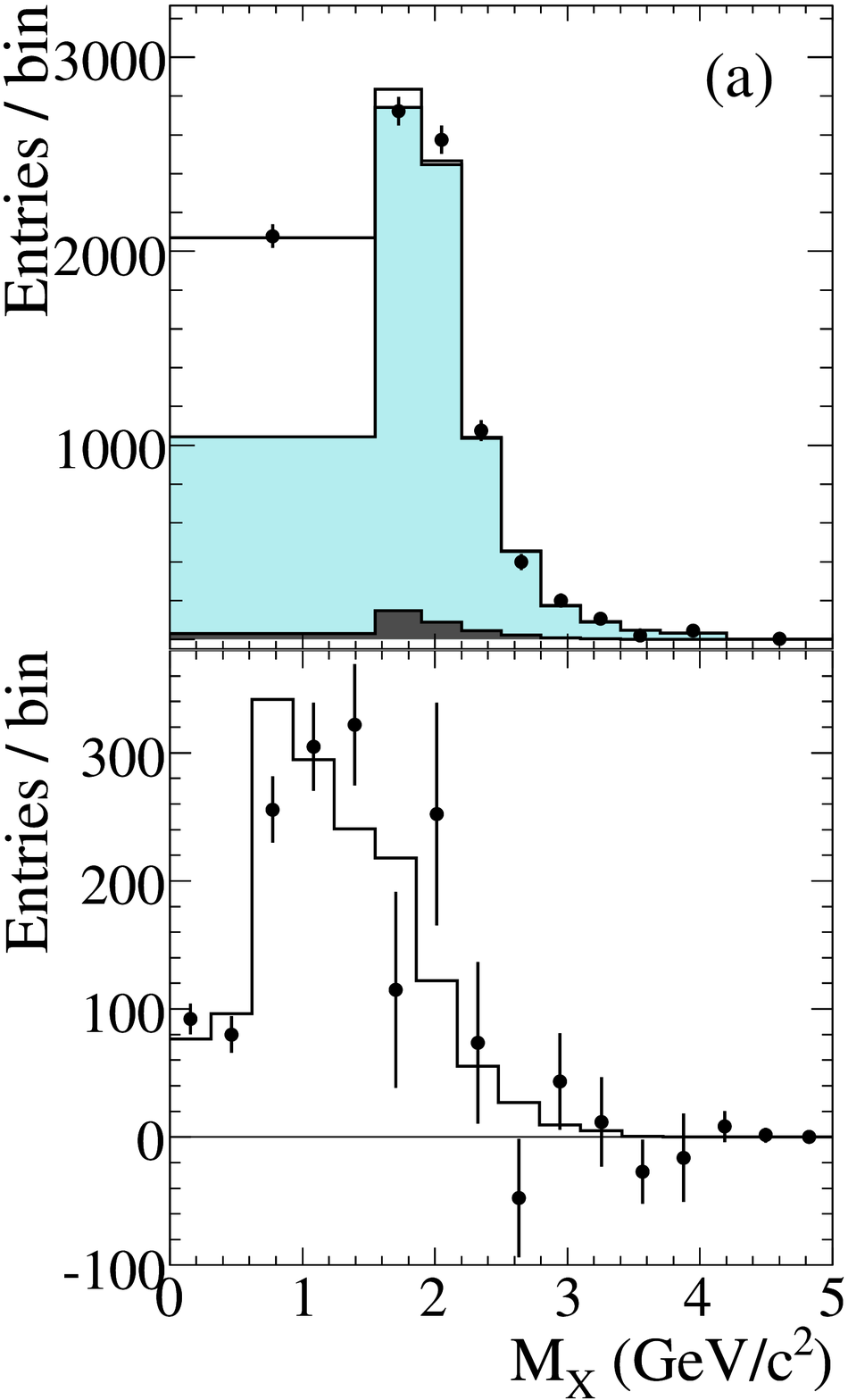}
      \includegraphics[width=0.245\textwidth,totalheight=7cm]{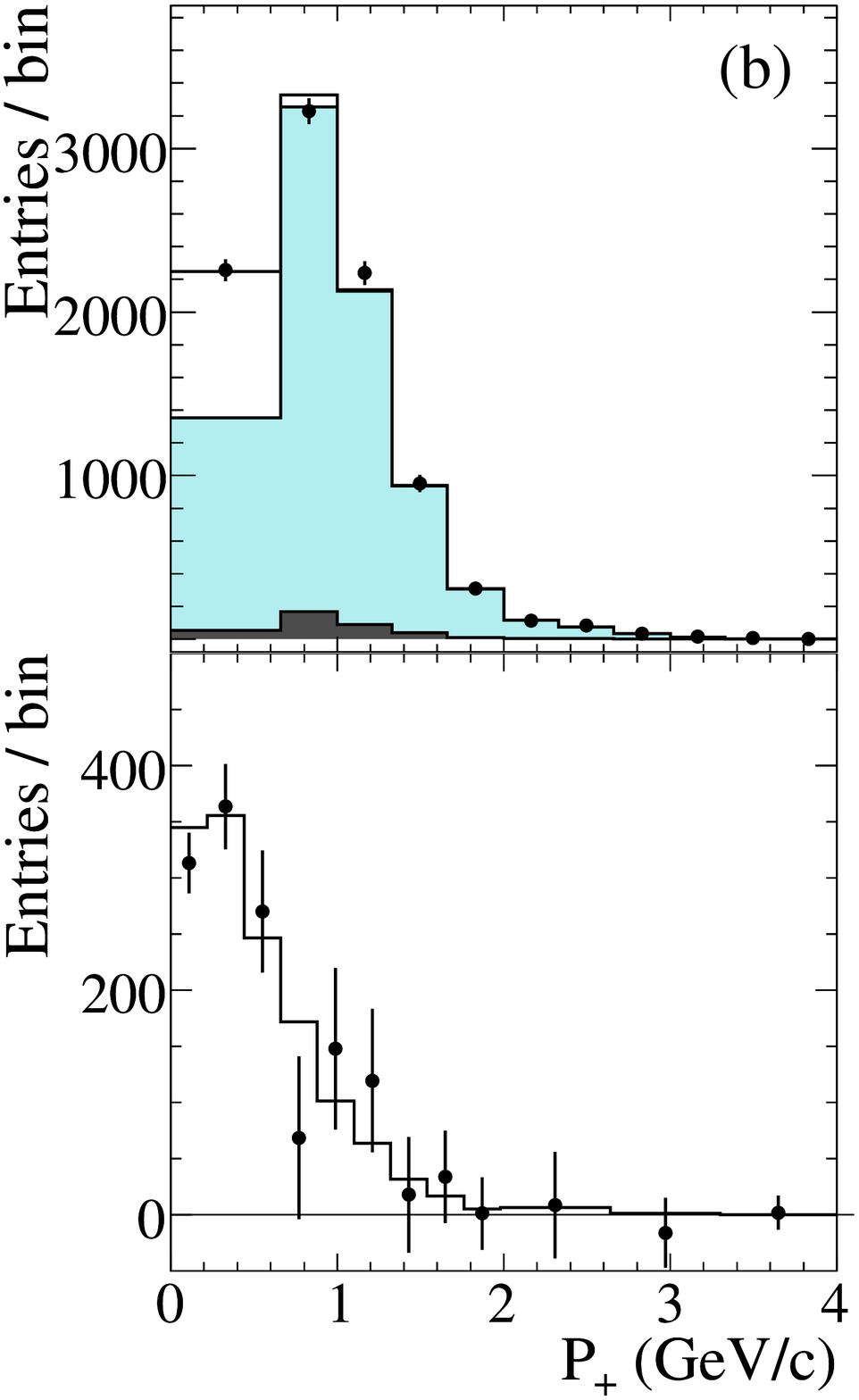}
      \includegraphics[width=0.245\textwidth,totalheight=7cm]{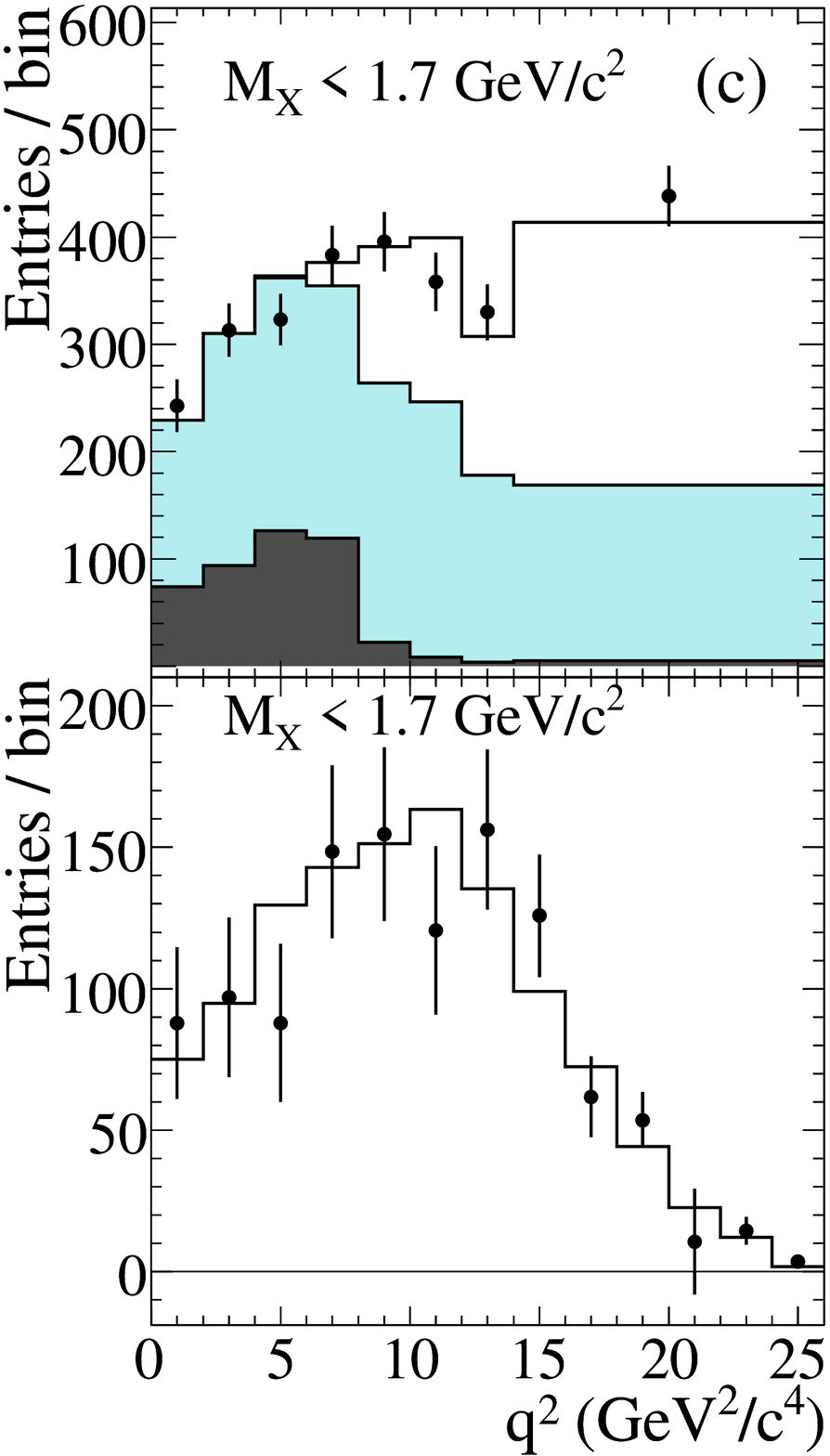}
      \includegraphics[width=0.245\textwidth,totalheight=7cm]{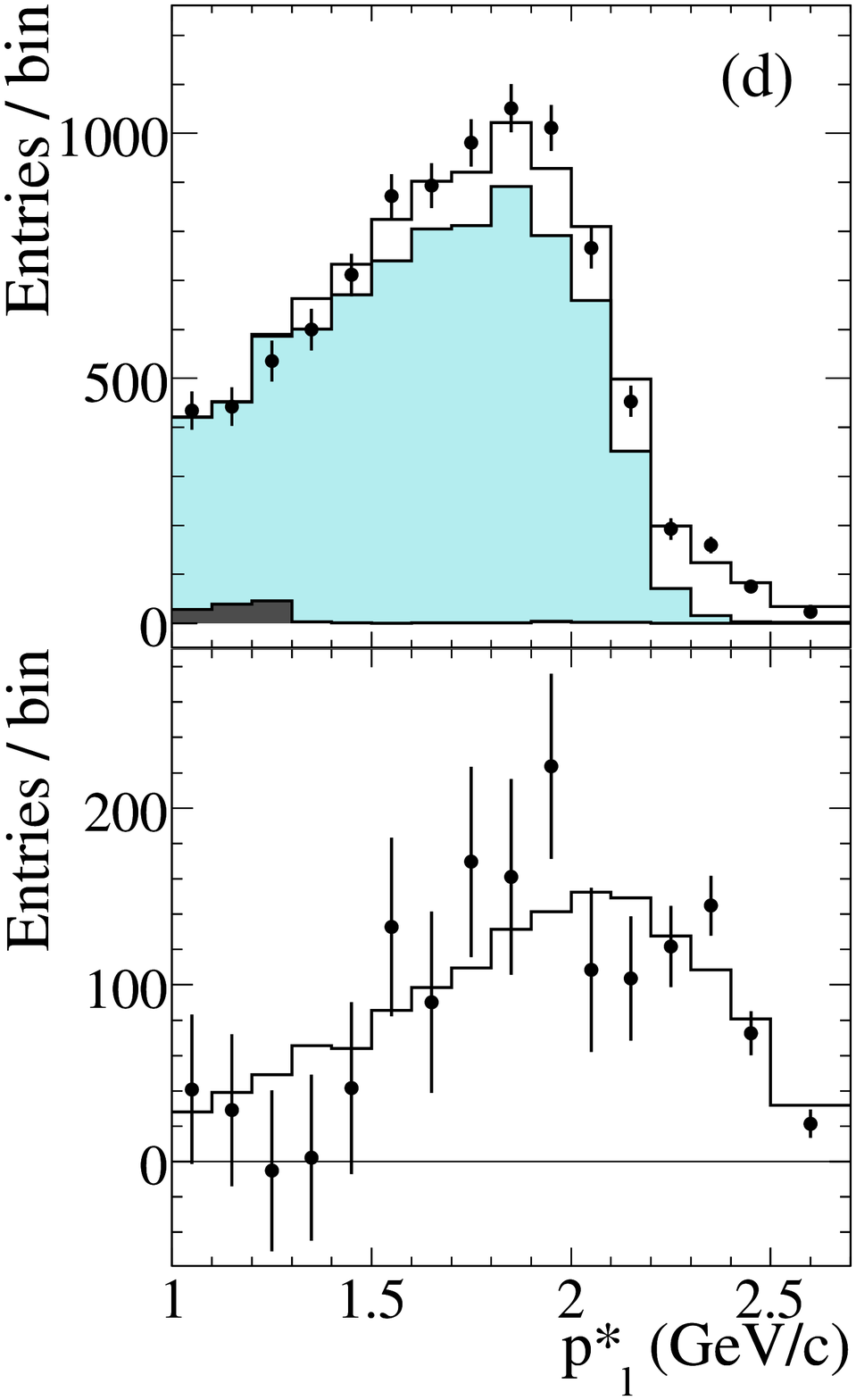}
      \caption{Upper row: The result of the MC fit to data (points) using the $\mx$ (a), $\Pplus$ (b),
        $q^2$ with $\mx<1.7~\gevcc$ (c) and $\Pl$ (d) spectra.
        The MC contributions correspond to $\Bxulnu$
        decays generated inside signal region (no shading), outside the signal region (dark shading),
        and the sum of $\Bxclnu$ and ``other'' backgrounds
        (light shading). Lower row: The same spectra after background subtraction using finer binning.}
    \label{fig:results}
   \end{centering}
   \end{figure*}

After these selection cuts we use various kinematic variables in differing phase space regions
to select the final signal sample.
This includes the lepton momentum ($\Pl$), the hadronic invariant mass ($\mx$),
the di-lepton invariant mass squared ($\Q$) and $\Pplus=E_{X}-|{\vec{P}}_{X}|$
\footnote{where $E_{X}$ and ${\vec{P}}_{X}$ are the energy and momentum of the hadronic $X$ system in the $B$ meson rest frame.}
We determine the number of $\Bxulnu$ signal
events and subsequently the PBF ($\Delta {\cal{B}}$) in the following optimised regions of phase space:
$M_X<1.55\gevcc$ and $M_X<1.70\gevcc$,
$P_+<0.66\gevc$,
$\Pl> 1.3 $\gev,
$M_X<1.70\gevcc$ and $~q^2>8.0~\mathrm{Gev^2/c^4}$,
and $M_X-q^2$ full phase space
using a $\chi^2$ minimisation fit of MC to data (for the last two cases a 2-dimensional fit is performed).
It is worth mentioning that we normalize the PBF to the total semileptonic branching fraction,
which reduces certain systematic uncertainies.
We find that the quality of the fit in regions dominated by the charm background (the high $\mx$ region for example) is poor.
So we rescale the contribution in MC from semileptonic decays into P-wave D mesons, which improves the
quality of the fit in these regions without affecting the final signal yield greatly.
The results are shown in Fig.~\ref{fig:results}.

\vspace{-0.35in}
\section{$\Vub$ results}
To translate the measured PBFs into $\Vub$ we use calculations from Neubert and
collaborators (BLNP) \cite{Lange:2005yw},
Andersen and Gardi (DGE) \cite{Andersen:2005mj},
Aglietti and collaborators (ADFR) ~\cite{Aglietti},
and Gambino and collaborators (GGOU) ~\cite{GGOU}.
Generally, we relate $\Vub$ to $\Delta \BR( \Bxulnu)$ with

\begin{equation}
|V_{ub}|  = \sqrt{\frac{\Delta \BR( \Bxulnu)}{\tau_B \cdot \Delta\Gamma_{theory} }}\\\nonumber ,
\end{equation}

\noindent
where $\tau_B = 1.573 \times 10^{-12}$~s is the lifetime of the $B$ meson \cite{PDG2008},
and $\Delta\Gamma_{theory}$ is the theoretical $\Bxulnu$
width according to the applied cuts. $\Delta\Gamma_{theory}$ varies from the calculation used,
and the experimental region. For this measurement we adopt and propagate the
uncertainty on this quantity as assessed by the authors.
We quote an estimate for $\Vub$ using the most inclusive measurement.
Namely, the the two-dimensional fit to the $\mx-\Q$ plane with no cuts other than $\Pl > 1.0 $\gev.
So, using $\Delta {\cal{B}} =(1.80 \pm 0.13_{\rm stat.} \pm 0.15_{\rm sys.}) \times 10^{-3}$
and calculating the arithmetic mean average of the four theoretical calculations
we find that \Vub = $(4.31 \pm 0.35) \times 10^{-3}$.

\section{Conclusion}

In summary, we present measurements of PBFs for the inclusive
charmless semileptonic decay of $\Bxulnu$,
in various regions of phase space. We require one of the of the $B$ mesons in the event to
be fully reconstructed in a hadronic decay mode, and the other $B$ to decay semileptonically into either
an electron or a muon. Using the analysis based on the two-dimensional fit on $\mx-\Q$
with no cut other than $\Pl> 1.0 $\gev,
we obtain \Vub = $(4.31 \pm 0.35) \times 10^{-3}$ as the most precise result. The total uncertainty is around
$8\%$, which is comparable to a similar study performed by Belle \cite{Belle_multivariate},
which uses a multivariate discriminant to reduce the background.
Furthermore, we place a limit on the potential contribution from the
Weak Annihilation of 9\% at 90\% C.L.

\end{document}